\documentclass{article}
\usepackage{ijcai25}

\usepackage{times}
\usepackage{soul}
\usepackage{url}
\usepackage[hidelinks]{hyperref}
\usepackage[utf8]{inputenc}
\usepackage[small]{caption}
\usepackage{graphicx}
\usepackage{amsmath}
\usepackage{amsthm}
\usepackage{booktabs}
\usepackage[switch]{lineno}
\usepackage{subcaption}
\usepackage{amssymb}
\usepackage{bm,amsfonts}
\usepackage[ruled]{algorithm2e}
\usepackage{multirow}
\usepackage{mathrsfs}
\usepackage{array,epsfig,fancyheadings,rotating}
\usepackage{color}
\usepackage[inline]{enumitem}

\newcommand{\bmu}{\boldsymbol{\mu}}
\newcommand{\bSigma}{\boldsymbol{\Sigma}}

\newcommand{\bphi}{\boldsymbol{\phi}}
\newcommand{\btheta}{\boldsymbol{\theta}}
\newcommand{\bz}{\boldsymbol{\mathrm{z}}}
\newcommand{\bgamma}{\boldsymbol{\gamma}}

\newcommand{\bO}{\boldsymbol{\mathrm{O}}}

\newcommand{\bp}{\boldsymbol{\mathrm{p}}}

\newcommand{\bH}{\boldsymbol{\mathrm{H}}}

\newcommand{\bQ}{\boldsymbol{\mathrm{Q}}}
\newcommand{\bK}{\boldsymbol{\mathrm{K}}}
\newcommand{\bV}{\boldsymbol{\mathrm{V}}}
\newcommand{\bM}{\boldsymbol{\mathrm{M}}}

\newcommand{\bXi}{\boldsymbol{\Xi}}
\newcommand{\bOmega}{\boldsymbol{\Omega}}

\newcommand{\bzeta}{\boldsymbol{\zeta}}
\newcommand{\mysmall}{\fontsize{8pt}{7pt}\selectfont}

\title{Utilizing Effective Dynamic Graph Learning to Shield Financial Stability from Risk Propagation}

 \author{
 Guanyuan Yu$^1$\and
 Qing Li$^1$\and
 Yu Zhao$^{1}$\and
 Jun Wang$^1$\and
 YiJun Chen$^1$\And
 Shaolei Chen$^2$\\
 \affiliations
 $^1$Southwestern University of Finance and Economics\\
 $^2$Sichuan XWBank\\
 \emails
 \{yuguanyuan, liq\_t, zhaoyu, wangjun, chenyijun\}@swufe.edu.cn,
 chenshaolei@xwbank.com
 }

\begin{document}

\maketitle
\begin{abstract}

    Financial risks can propagate across both tightly coupled temporal and spatial dimensions, posing significant threats to financial stability. Moreover, risks embedded in unlabeled data are often difficult to detect. To address these challenges, we introduce \textbf{GraphShield}, a novel approach with three key innovations:
    \textbf{Enhanced Cross-Domain Information Learning}: We propose a dynamic graph learning module to improve information learning across temporal and spatial domains.
    \textbf{Advanced Risk Recognition}: By leveraging the clustering characteristics of risks, we construct a risk recognizing module to enhance the identification of hidden threats.
    \textbf{Risk Propagation Visualization}: We provide a visualization tool for 
    quantifying and validating nodes that trigger widespread cascading risks.
    Extensive experiments on two real-world and two open-source datasets demonstrate the robust performance of our framework. Our approach represents a significant advancement in leveraging artificial intelligence to enhance financial stability, offering a powerful solution to mitigate the spread of risks within financial networks.
\end{abstract}

\section{Introduction}
In financial markets like the networked-guarantee loan market, 
entities such as small and medium-sized enterprises (SMEs) are integrated into the same ecosystem. 
Within this network, the default risk of one SME can be influenced by the financial health and behavior of its peers. 
This interconnectedness leads to a phenomenon known as financial risk propagation~\cite{ali2020shared}. 
Without adequate management, this can lead to widespread cascading risks, potentially destabilizing the entire financial system~\cite{eisenberg2001systemic,billio2012econometric,elliott2014financial}.

\begin{figure}[htbp]
	\setlength{\abovecaptionskip}{0pt}
	\setlength{\belowcaptionskip}{0pt}
	\centering
	\includegraphics[width=7cm]{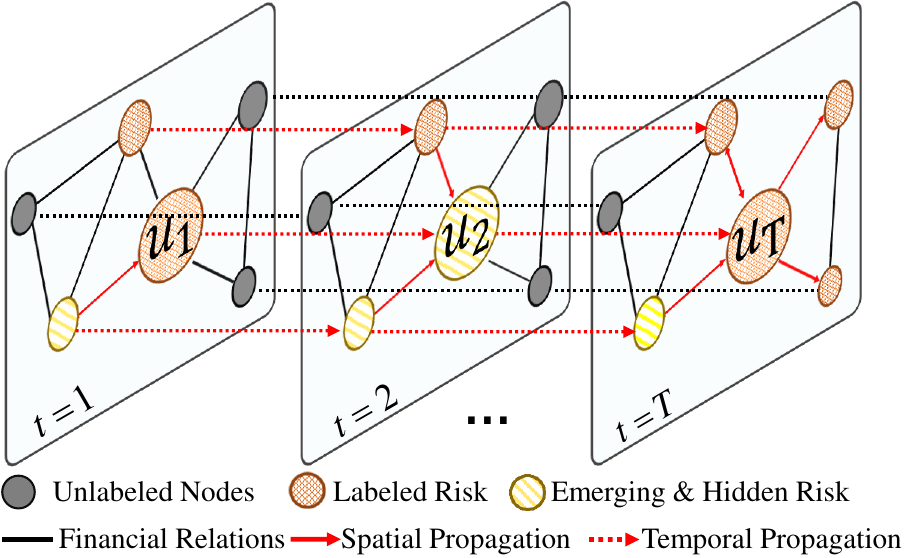}
    \caption{\textbf{An Example to Illustrate Financial Risk Propagation.} (a) Financial risks are intricately interconnected across spatial and temporal domains. For instance, the risk status of a node (e.g., $u_t$) is shaped by both its immediate neighbors and its previous state ($u_{t-1}$). (b) Risk nodes (e.g., $u_2$) hidden in unlabelled data drive the propagation of risk. Its risk status influences both its immediate neighbors and its future vulnerabilities. Failing to effectively identify these nodes can result in uncontrolled risk propagation.
    (c) Risk samples frequently display clustering patterns in both temporal and spatial dimensions. For example, if certain nodes are identified as risky, their neighboring nodes and previous states might also pose potential risks.
    }
	\label{fig:risk analysis}
\end{figure}
    As shown in Fig.~\ref{fig:risk analysis}, financial risks can propagate through both temporal and spatial domains, which are tightly coupled. For example, the risk status of a node (e.g., $u_t$) is influenced by both its immediate neighbors and its prior state ($u_{t-1}$). 
    This interdependence complicates the understanding of their propagation mechanisms.
    %
    Recent studies have utilized dynamic graph neural networks to represent the financial system as a graph, where entities are nodes and their interconnections are edges~\cite{cheng2022regulating}.
    %
    Such models typically capture structural features using graph neural networks, such as GCNs and GATs. They then proceed to learn temporal features through time-series models, including gated recurrent units, temporal attention layers, and iTransformer.
    Representative models include AddGraph~\cite{zheng2019addgraph}, TRACER~\cite{cheng2020contagious}, StrGNN~\cite{cai2021structural}, and RisQNet~\cite{ijcai2024p817}.
    %
    It has been demonstrated that processing spatial and temporal information separately within such hybrid frameworks can lead to information loss across temporal and spatial domains, resulting in suboptimal outcomes~\cite{liu2021anomaly}.

    %
    As the scope and nature of real-world financial businesses continue to evolve, the patterns and forms of financial risks dynamically change~\cite{hanley2019dynamic}. 
    This evolution gives rise to new risks that may emerge stealthily and remain undetected or unlabeled, posing significant challenges to risk management strategies.
    As illustrated in Fig.~\ref{fig:risk analysis}, identifying some risk nodes hidden within the unlabeled data is particularly challenging yet crucial for preventing risk propagation. For instance, the risk status of $u_2$ impacts both its immediate neighbors and its future vulnerabilities.
    Failure to effectively identify and address these hidden risks can severely hinder efforts to control their spread.
    %
    %
    %

In this study, we introduce \textbf{GraphShield}, a novel and effective dynamic graph learning approach designed to protect financial stability from the propagation of risks. This approach can achieve the following three main functionalities:
(a) To enhance information learning across spatial and temporal domains, which are tightly coupled, we integrate both spatial and temporal operations simultaneously into a single layer of the dynamic learning module.
This integration structure ensures that it captures the structural information of nodes while concurrently learning their temporal information.
%
(b) To enhance the identification of hidden risks, we extend beyond the use of risk labels by exploiting the clustering tendencies of risk samples. These samples frequently group together, as visually demonstrated in Fig.~\ref{fig:risk analysis} and empirically validated in the study~\cite{ijcai2024p817}. We hypothesize that risk samples follow a Gaussian mixture distribution and employ a fully-connected neural network to construct the risk recognizing module, which can reduce the over-reliance on labels.
(c) Furthermore, we offer a financial risk propagation visualization analysis tool capable of quantifying and validating the impact between risks. This tool aids in pinpointing and quantifying key factors and entities that trigger widespread cascading risks, exploring the interactions among various factors and entities that generate risk, and devising strategies to effectively mitigate or manage these risks.
\begin{figure*}[htbp]
	\setlength{\abovecaptionskip}{0pt}
	\setlength{\belowcaptionskip}{0pt}
	\centering
	\includegraphics[width=17.8cm]{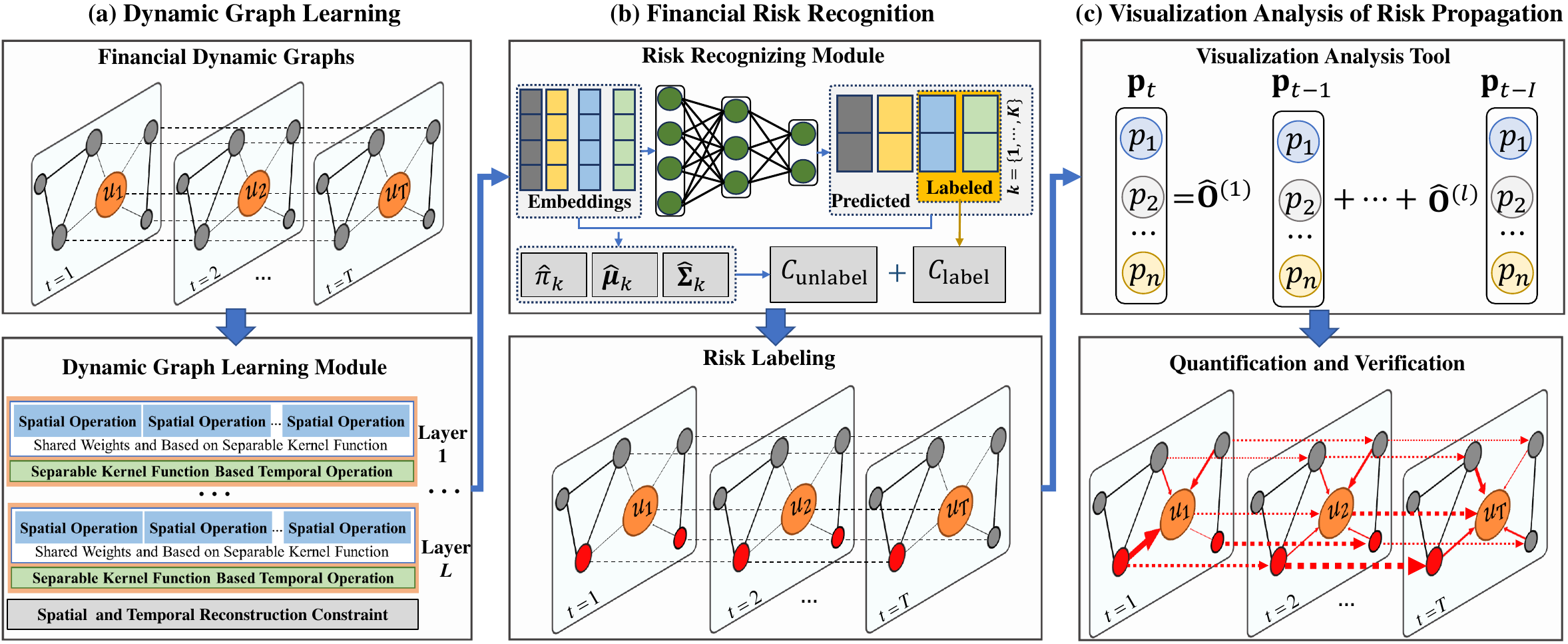}
	\caption{\textbf{Overall Architecture of Our Proposed GraphShield Framework.} 
	(a) We construct the dynamic graph learning module integrating spatial and temporal operations within each layer.
	The sandwich-style stacking structure ensures thorough learning of spatial and temporal information.
	(b) We represent each risk sample as originating from a Gaussian mixture distribution and employ a fully-connected neural network to enhance the hidden risk recognition.
	(c) We offer a financial risk propagation visualization analysis tool to quantifying and validating the impact effects between risks.
	} 
	\label{fig:overall}
\end{figure*}
%
%
In summary, our study presents the following three unique contributions.

    
    (a) We propose a novel dynamic graph learning module to enhance information learning across spatial and temporal domains by 
    integrating both spatial and temporal operations into a single layer. 
    Besides, We are pioneering research into identifying hidden yet critical risk entities in the financial risk propagation process by leveraging their  clustering characteristics. 
    
   (b) We conduct a rigorous evaluation of our proposed approach by comparing it with existing benchmarks across various datasets, achieving state-of-the-art performance. Additionally, beyond mere risk identification, we offer an in-depth visualization analysis of financial risk propagation and demonstrate that our approach can prevent significant financial losses.
   
   (c) Our method represents an advancement in leveraging artificial intelligence to enhance financial stability. It offers a strong framework to mitigate the spread of risk throughout financial networks. By doing so, it strengthens financial stability, promotes economic growth, and aligns with sustainable development goals.
    
\section{Related Work}


Recent studies have utilized machine learning and deep learning technologies to capture and analyze complex data patterns, offering new perspectives and methods for risk assessment and management. 
%
For example, TRACER~\cite{cheng2020contagious}, iConReg~\cite{cheng2022regulating}, SCRPF~\cite{cheng2023critical}, RisQNet~\cite{ijcai2024p817}
utilize graphs to depict the loan-guarantee relationships among small and medium-sized enterprises in the networked loan market, and construct effective deep graph neural networks to identify and curb the propagation of financial defaults.

In this paper, acknowledging the dynamic propagation of financial risks across spatial and temporal dimensions, and their evolving patterns, we introduce a novel and effective dynamic graph learning model designed for the recognition and analysis of financial risks.
\section{Preliminary}

Financial dynamic graphs can be represented as $\mathcal{G} = \{\mathcal{G}^{\mathsf{S}}, \mathcal{G}^{\mathsf{T}}\}$. The spatial domain $\mathcal{G}^{\mathsf{S}}$ comprises a series of directed heterogeneous graphs $\mathcal{G}^{\mathsf{S}}_t = \{\mathcal{A}_t, \mathcal{E}_t, \mathcal{B}_t, \mathcal{R}_t\}$ observed at discrete times $t = \{1, \dots, T\}$. Here, $\mathcal{A}_t$ and $\mathcal{E}_t$ denote the sets of nodes and edges at time $t$, respectively. Each node $u \in \mathcal{A}_t$ and each edge $e \in \mathcal{E}_t$ are linked to their respective types through the type mapping functions $\eta(\cdot)$ and $\varphi(\cdot)$, associating nodes with types $\eta(u) \in \mathcal{B}_t$ and edges with types $\varphi(e) \in \mathcal{R}_t$.
The temporal domain $\mathcal{G}^{\mathsf{T}}$ is defined as $\{\mathcal{G}_u^{\mathsf{T}}: u \in \mathcal{A}\}$, where $\mathcal{A}$ includes all node types. For each node $u$, $\mathcal{G}^{\mathsf{T}}_u$ is a directed homogeneous graph that captures the evolution of node $u$ across the time points $t = \{1, \dots, T\}$. This graph consists of nodes $\mathcal{A}_u = \{u_1, u_2, \ldots, u_T\}$ and edges $\mathcal{E}_u = \{e(u_i, u_j): e(u_i, u_j) = 1 \text{ if } j = i+1 \text{ and } e(u_i, u_j) = 0 \text{ otherwise}; i, j = 1, \ldots, T\}$.



\section{Our Proposed GraphShield Approach}
As illustrated in Fig.~\ref{fig:overall}, our proposed GraphShield approach can achieve the following three functionalities: (a) dynamic graph learning, (b) financial risk recognition, and (c) visualization analysis of risk propagation.

\subsection{Dynamic Graph Learning Module}

To effectively learn from dynamic graphs, neural network models must integrate both spatial structure and temporal dynamics. Typically, these two types of information are intertwined and must be processed simultaneously to enhance financial risk detection. 
%
A critical design consideration for dynamic graph encoders is \textit{how to simultaneously account for spatial and temporal information}.  Existing models involve using hybrid networks that combine spatial and temporal modules. These modules independently capture spatial and temporal data. 
For example, in the StrGNN model~\cite{cai2021structural}, a Graph Convolutional Network (GCN) acts as the spatial module, while a Gated Recurrent Unit (GRU) processes the GCN outputs across different timestamps to handle temporal dynamics. However, this separation can lead to the loss of information across spatial and temporal domains, resulting in suboptimal performance~\cite{liu2021anomaly}.
%
%
To address this issue, we introduce a novel dynamic graph learning module that interleaves spatial and temporal operations in a sandwich-like structure. Additionally, we implement both spatial and temporal operations using a multi-head attention mechanism enhanced by a separable kernel function, effectively reducing the time complexity from quadratic to linear.

\subsubsection{Separable Kernel Function Based Spatial Operation}

Here, we apply the separable kernel function based multi-head attention mechanism to construct spatial operations on a graph. Specifically, given a graph $\mathcal{G}_t^{\mathsf{S}}$ at timestamp $t$, let $u_t$ represent the target node. Let $v \in \mathsf{N}(u_t)$ denote the neighbors of $u_t$.
In the $h$-th head of the multi-head attention mechanism, we apply a node type-specific linear transformation $\mathrm{Q\text{-}Linear}^{(h)}_{\eta(u_t)}(\cdot)$ to the target features $\bH^{(l-1)}_{u_t}$, converting them into a query matrix $\bQ$. 
Similarly, for each neighbor $v \in \mathsf{N}(u_t)$, we employ $\mathrm{K\text{-}Linear}^{(h)}_{\eta(v)}(\cdot)$ and $\mathrm{V\text{-}Linear}^{(h)}_{\eta(v)}(\cdot)$ to map $\bH^{(l-1)}_v$ into key and value matrices $\bK_v$ and $\bV_v $, respectively. 
These transformations are tailored to the node types, enhancing the ability of our model to capture and utilize the structural and feature diversity within the graph.

\begin{equation}
    \mysmall
    \begin{aligned}
      &\bQ=\mathrm{Q\text{-}Linear}^{(h)}_{\eta(u_t)}(\bH^{(l-1)}_{u_t})\in \mathbb{R}^{N_{u_t} \times \frac{d}{H}},\\
      &\bK=\mathrm{K\text{-}Linear}^{(h)}_{\eta(v)}(\bH^{(l-1)}_v)\in \mathbb{R}^{N_v \times \frac{d}{H}},\\
      &\bV=\mathrm{V\text{-}Linear}^{(h)}_{\eta(v)}(\bH^{(l-1)}_v)\in \mathbb{R}^{N_v \times \frac{d}{H}},\\
      & h=\{1, 2, \dots, H\}.
    \end{aligned}
\end{equation}

Notably, the canonical attention mechanism can be formulated as ${\bQ\bK^\top}\bV/{\sqrt{d/H}}$, which exhibits quadratic time complexity. To achieve a linear time complexity, we employ attention mechanisms based on separable kernel functions.
Specifically, we calculate the $i$-th row of the weighted message head $\bM^{(h)}$
via leveraging the following equation,
\begin{equation}
   \mysmall
   \begin{aligned}
    \bM^{(h)}_i &= \frac{\sum_{j=1}^{N_v}\mathrm{sim}(\bQ_i,\bK_j)\bV_j}{\sum_{j=1}^{N_v}\mathrm{sim}(\bQ_i,\bK_j)}
    = \frac{\sum_{j=1}^{N_v}\phi(\bQ_i)\phi(\bK_j)^\top\bV_j}{\sum_{j=1}^{N_v}\phi(\bQ_i)\phi(\bK_j)^\top}\\ &=\frac{\phi(\bQ_i)\sum_{j=1}^{N_v}\phi(\bK_j)^\top\bV_j}{\phi(\bQ_i)\sum_{j=1}^{N_v}\phi(\bK_j)^\top}.
\end{aligned} 
\end{equation}

In the above equation, 
the kernel function is defined as $\phi(x) = \mathrm{ELU}(x) + 1$. Given that both $\sum_{j=1}^{N_v}\phi(\bK_j)^\top\bV_j$ and $\sum_{j=1}^{N_v}\phi(\bK_j)^\top$ can be precomputed, cached, and reused, the complexity of calculating $\bM^{(h)}$ can be reduced from quadratic to linear. This optimization significantly enhances the efficiency of the process.

Then, the updated node representation $\bH^{(l)}_{u_t}$ is computed as $\tau_1\bH^{(l-1)}_{u_t} + (1-\tau_1)\bigoplus_{h=1}^H\bM^{(h)}$. Here, $\bigoplus_{h=1}^H\bM^{(h)}$ denotes the concatenation of $H$ message heads, and $\tau_1$, a trainable parameter, lies within the interval (0,1). 

\subsubsection{Separable Kernel Function Based Temporal Operation}


To construct temporal operations, we employ $H$ heads of multi-head attention mechanisms based on separable kernel functions. 
Initially, we incorporate rotary position encoding~\cite{su2024roformer} prior to the temporal operation to mark the temporal order and relevance of the node sequence $\{u_t\}_{t=1}^{T}$. 
For $\bH^{(l)}_{u_t}$ at timestamp $t$, the rotated position embedding $\widetilde{\mathbf{H}}_{u_t}$ is computed as follows,

\begin{equation}
    \mysmall
    \begin{aligned}
  \widetilde{\mathbf{H}}_{u_t,2i} &= \mathbf{H}^{(l)}_{u_t,2i}\cdot \cos\left(\frac{t}{10000^{2i/d}}\right) \\ & + \mathbf{H}^{(l)}_{u_t,2i+1} \cdot \sin\left(\frac{t}{10000^{2i/d}}\right),
    \end{aligned}
\end{equation}

\begin{equation}
    \mysmall
    \begin{aligned}
  \widetilde{\mathbf{H}}_{u_t,2i+1} & = \mathbf{H}^{(l)}_{u_t,2i+1} \cdot \cos\left(\frac{t}{10000^{2i/d}}\right)\\ &-\mathbf{H}^{(l)}_{u_t,2i} \cdot \sin\left(\frac{t}{10000^{2i/d}}\right),
    \end{aligned}
\end{equation}
where $t=\{1, 2, \dots, T\}$ and $i=\{1, 2, \dots, d\}$. 
%
At each layer, the input vectors are multiplied by their corresponding rotation vectors. In different layers, these input vectors undergo various rotational encodings. Consequently, the shallower layers primarily focus on information from neighboring positions, while the deeper layers concentrate on information from more distant locations.
This hierarchical rotation allows the temporal operation to capture positional information from multiple perspectives, thereby enhancing its understanding of the global structure and context of the sequential data.

In the $h$-th attention head, we derive $\bM^{(h)}$ through the following linear transformation,
\begin{equation}
    \mysmall
    \begin{aligned}
  &\bQ=\mathrm{Q\text{-}Linear}^{(h)}(\widetilde{\bH}_{u}) \in \mathbb{R}^{T \times \frac{d}{H}},\\
    	&\bK=\mathrm{K\text{-}Linear}^{(h)}(\widetilde{\bH}_{u}) \in \mathbb{R}^{T \times \frac{d}{H}},\\ 
    	&\bV=\mathrm{V\text{-}Linear}^{(h)}(\widetilde{\bH}_{u}) \in \mathbb{R}^{T \times \frac{d}{H}},\\
    	&\bM^{(h)}_i=\frac{\phi(\bQ_i)\sum_{j=1}^{T}\phi(\bK_j)^\top\bV_j}{\phi(\bQ_i)\sum_{j=1}^{T}\phi(\bK_j)^\top}.
    \end{aligned}
\end{equation}


In the equation above, the kernel function $\phi(x) = \mathrm{ELU}(x) + 1$. The terms $\sum_{j=1}^{T}\phi(\bK_j)^\top\bV_j$ and $\sum_{j=1}^{T}\phi(\bK_j)^\top$ can be precomputed, cached, and reused, enabling linear computational complexity. Finally, the update for $\bH^{(l+1)}_u$ is given by $\tau_2\bH^{(l)}_u + (1-\tau_2) \bigoplus_{h=1}^H \bM^{(h)}$, where $\tau_2 \in (0,1)$ is a trainable parameter.




\subsubsection{Graph Reconstruction Constraint}
\begin{equation*}
    \mysmall
    \begin{aligned}
    C_\mathrm{rec}(\widehat{\mathcal{G}}, \mathcal{G}) = \underbrace{\sum_{t=1}^{T}|D(\widehat{\mathcal{G}}_t^\mathsf{S}) - D(\mathcal{G}^\mathsf{S}_t)|_F^2}_{\text{Spatial reconstruction}}
    + \underbrace{\sum_{u \in \mathcal{A}}|D(\widehat{\mathcal{G}}^\mathsf{T}_u) - D(\mathcal{G}^\mathsf{T}_u)|_F^2}_{\text{Temporal reconstruction}},
    \end{aligned}
\end{equation*}
where 
$\widehat{\mathcal{G}} = \{\widehat{\mathcal{G}}^{\mathsf{S}}, \widehat{\mathcal{G}}^\mathsf{T}\}$ represents the predicted graph, and $D(\cdot)$ computes the adjacency matrix. Minimizing $C_\mathrm{rec}(\widehat{\mathcal{G}}, \mathcal{G})$ optimizes the model parameters.

\subsection{Risk Recognizing Module}
To enhance the identification of hidden risks, we move beyond traditional reliance on risk labels. Instead, we leverage the inherent clustering tendencies of risk samples, which naturally group together. This is visually demonstrated in Fig.~\ref{fig:risk analysis} and empirically validated in the study~\cite{ijcai2024p817}.
Here, let $\mathcal{Z}=\{\bz_1,\ldots,\bz_N\}$ denote the feature vectors learned by the dynamic graph learning module, as discussed in the previous section, where $N=T\times |\mathcal{A}|$. 
Intuitively, $\mathcal{Z}$ can be categorized into $K$ distinct groups, denoted as $\{R_k\}_{k=1}^K$. To approximate the distribution of $\mathcal{Z}$, we propose using a Gaussian mixture model (GMM) defined as $\mathcal{P}_{\rm data}(\bz|\btheta)=\sum_{k=1}^K\pi_k\mathcal{N}(\bz|\bmu_k, \bSigma_k)$. Here, $\btheta=\{\pi_k, \bmu_k, \bSigma_k; k=1,\ldots,K\}$ represents the model parameters, subject to the constraint $\sum_{k=1}^K\pi_k=1$.
To estimate $\btheta$, we construct the following fully-connected networks,
\begin{equation}
    \mysmall
    \begin{aligned}
  \mathbf{h}^{(l)} = \text{BatchNorm}\left(\text{ReLU}(\mathbf{W}^{(l)}\mathbf{h}^{(l-1)} + \mathbf{b}^{(l)})\right),
    \end{aligned}
\end{equation}
where $l=\{1, 2, \dots, L\}$ and $\mathbf{h}^{(0)} = \bz_i$. The final output of such fully-connected networks is as follows,

\begin{equation}
    \mysmall
    \begin{aligned}
  &\bgamma_{i} = [\gamma_{i,1},\ldots,\gamma_{i,K}]^\top = \text{Softmax}(\mathbf{h}^{(L)}) \in \mathbb{R}^K.
    \end{aligned}
\end{equation}

Next, we calculate the estimated expectation $\widehat{\bmu}_k$, component probability $\hat{\pi}_k$, and covariance $\widehat{\bSigma}_k$ for the $k$-th data group,
\begin{equation}
    \mysmall
    \begin{aligned}
  \widehat{\bmu}_k &= \frac{\sum_{i=1}^{N}\gamma_{i,k}\bz_i}{\sum_{i=1}^{N}\gamma_{i,k}},~~~\hat{\pi}_k = \frac{1}{N}\sum_{i=1}^{N}\gamma_{i,k},\\
  \widehat{\bSigma}_k &= \frac{\sum_{i=1}^{N}\gamma_{i,k}(\bz_i - \widehat{\bmu}_k)(\bz_i - \widehat{\bmu}_k)^\top}{\sum_{i=1}^{N}\gamma_{i,k}}.
    \end{aligned}
\end{equation}

To optimize the network parameters, we introduce the following loss function,
\begin{equation}
    \mysmall
	\begin{aligned}
	\mathcal{L}(\mathcal{Z}; \bphi)=&\underbrace{-\frac{1-\tau_3}{N}\sum_{i=1}^{N}\log\left[\sum_{k=1}^{K}\hat{\pi}_k
		\mathcal{N}\left(\bz_i~|~\widehat{\bmu}_k,\widehat{\bSigma}_k\right)\right]}_{C_{\mathrm{unlabel}}}\\
		&\underbrace{-\tau_3\sum_{i=1}^{N}\sum_{k=1}^{K}\mathbb{I}_{\mathcal{R}_k\cap\mathcal{D}_l}(\bz_i)\log(\gamma_{i,k})}_{C_\mathrm{label}}+C_\mathrm{rec}\,,
	\end{aligned}
\end{equation}
where $\mathcal{D}_l$ and $\mathcal{D}_u$ denote the labeled and unlabeled datasets, respectively, with $|\mathcal{D}_l| \ll |\mathcal{D}_u|$ and $|\mathcal{D}_l| + |\mathcal{D}_u| = N$. The parameter $\tau_3$ is used to adjust the weighting of the components in the objective function. The indicator function $\mathbb{I}_{\mathcal{R}_k \cap \mathcal{D}_l}(\bz_i)$ signifies that $\bz_i$ is classified into the $k$-th group $\mathcal{R}_k$ and has a ground-truth label.
$C_{\mathrm{unlabel}}$ can model the clustering tendencies of risks, thereby enhancing the identification of unlabeled risks.

\subsection{Visualization Analysis Tool}
Based on the risk recognition,  
this subsection provide a robust visualization analysis tool to estimate and validate the impact effects within the financial risk propagation process. Let $\bp_t = [p_{1,t}, \ldots, p_{n,t}]^{\top}$ denote the probabilities of risks at $n$ nodes explored at time $t$.
A visualization analysis tool with a lag of $I$ is described by the following equations,
\begin{equation}
    \mysmall
    \begin{aligned}
       	&\bp_t=\sum_{\ell=1}^I\bO^{(\ell)}\bp_{t-\ell}+\bzeta_t,\\
	& \text{temporal constraint}:\\
    &\bO^{(\ell)}_{i, j} 
    \begin{cases}
    \neq 0 & \text{if edge}~e(p_{i,t-\ell}\to p_{j,t})\in \mathcal{E},\\
    = 0 & \text{otherwise}
    \end{cases} 
    \\
    & \text{where}~\ell=\{1,\ldots, I\}.\\
    & \text{spatial constraint}:\\
    &\bOmega_{i,j} 
    \begin{cases}
    \neq 0 & \text{if edge}~e(p_{i,t}\to p_{j,t})\in \mathcal{E},\\
    = 0 & \text{otherwise},
    \end{cases} \label{eq:spatial}\\
    & \text{where}~t=\{1,\ldots, T\}.
    \end{aligned}
\end{equation}

In the above equations,
$\bO^{(\ell)}$ is a $n \times n$ coefficient matrix, and $\bzeta_t$ is a $n$-dimensional white noise vector, distributed as $\bzeta_t \sim \mathcal{N}(\mathbf{0}, \bXi)$, where $\bXi$ is a $n \times n$ covariance matrix.
The element $\bO^{(\ell)}_{i,j}$ indicates that the risk probability $p_{i,t-\ell}$ Granger-causes $p_{j,t}$ if $\bO^{(\ell)}_{i,j} \neq 0$ \cite{eichler2005graphical}. 
Similarly, $\bOmega_{i,j}$, an element of the precision matrix $\bOmega = \bXi^{-1}$, signifies spatial dependence between $p_{i,t}$ and $p_{j,t}$ if $\bOmega_{i,j} \neq 0$ at a given timestamp $t$.

To estimate the coefficients $\bO = \{\bO^{(\ell)}\}_{\ell=1}^I$ and the precision matrix $\bOmega$, this study employs a penalized maximum likelihood estimation approach. This method improves the identification of model parameters by incorporating a penalty term, which increases the likelihood that certain elements within the coefficient and precision matrices are estimated as zero.
\begin{equation}
    \mysmall
    \begin{aligned}
  &\widehat{\bO}, \widehat{\bOmega}  = \arg\min_{\bO, \bOmega} \bigg\{
  g_1(\bO, \bOmega) + g_2(\bO, \bOmega) \bigg\},\\
  &g_1(\bO, \bOmega)=\sum_{t=1}^{T}\bzeta_t^{\top}\bOmega \bzeta_t -\frac{n}{2}\log |\bOmega|,\\
  &g_2(\bO, \bOmega) = \lambda_1 \sum_{i \neq j}^{n} |\bOmega_{i,j}| + \lambda_2 \sum_{\ell=1}^I \sum_{i,j = 1}^{n} |\bO^{(\ell)}_{i,j}|,
    \end{aligned}
\end{equation}
where $\lambda_1,\lambda_2\geq 0$ are regularization parameters that control the sparsity of $\bOmega$ and $\bO$, respectively.

To estimate Granger causality in the spatial domain, we utilize the partial contemporaneous correlation (PCC) \cite{dahlhaus2003causality}, which can eliminate the impact of other nodes on the correlation between $p_{i,t}$ and $p_{j,t}$. The PCC is defined as,
\begin{equation}
    \mysmall
    \begin{aligned}
       {\rm PCC}(p_{i,t}, p_{j, t}) = -{\widehat{\bOmega}_{i,j}}/{\sqrt{\widehat{\bOmega}_{i,i}\widehat{\bOmega}_{j,j}}}.
    \end{aligned}
\end{equation}

To measure Granger causality in the temporal domain, we employ the following partial directed correlations (PDC) \cite{dahlhaus2003causality,eichler2005graphical}, which removes the linear influence of other nodes on the correlation between $p_{i,t-\ell}$ and $p_{j,t}$. 
The PDC is given by,
\begin{equation}
    \mysmall
    \begin{aligned}
       {\rm PDC}(p_{i,t},p_{j,t-\ell})={\widehat{\bO}^{(\ell)}_{i,j}}/{\sqrt{z\widehat{\bXi}_{i,i}}},
    \end{aligned}
\end{equation}
where $z=\widehat{\bOmega}_{j,j}+\sum_{\delta=1}^{\ell-1}\sum_{\alpha,\beta=1}^{n}\widehat{\bO}^{(\delta)}_{\alpha,j}\widehat{\bOmega}_{\alpha,\beta}\widehat{\bO}^{(\delta)}_{\beta,j}$.
Finally, we employ a likelihood ratio test to assess the significance of each element in PCC and PDC.

\section{Risk Recognition Performance}
In this section, we validate the ability of our proposed GraphShield to identify risks on two real-world and two open-source datasets.

\subsection{Data Description}
We conduct experiments using four datasets from distinct financial domains: Bank-Partner, Shareholding, Bitcoin-OTC~\cite{kumar2016edge}, and Bitcoin-Alpha~\cite{kumar2018rev2}.
A summary of these datasets is provided in Table~\ref{tab:dataset}.
\begin{table}[h]
    \setlength{\tabcolsep}{0.9mm}
    \fontsize{7.5pt}{6pt}\selectfont
    \setlength{\abovecaptionskip}{0pt}
    \setlength{\belowcaptionskip}{0pt}
    \centering
    \begin{tabular}{c|c|c|c|c|c}
  \toprule
  Data & \# Nodes & \# Edges & Period & Freq. & \begin{tabular}[c]{@{}c@{}}Time Steps\\ (Train / Test)\end{tabular} \\
  \midrule
  Bank-Partner & 734K & 1.08M & 1/1/17-12/31/22 & Monthly & 48 / 24 \\
  Shareholding & 1.04M & 1.57M & 3/31/13-9/30/22 & Quarterly & 30 / 8 \\
  Bitcoin-OTC & 5.9K & 35.6K & 11/8/10-1/24/16 & Weekly & 95 / 42  \\
  Bitcoin-Alpha & 3.8K & 24.2K & 11/7/10-1/21/16 & Weekly & 95 / 41  \\ 
  \bottomrule
    \end{tabular}
    \caption{\textbf{Dataset Summary}.}
    \label{tab:dataset}
\end{table}

\textbf{Bank-Partner}:
This dataset is collected from an Internet commercial bank in Sichuan and includes consumer loan application records from January 2017 to December 2022. It covers approximately 10,500 bank partners, who are responsible for customer acquisition and loan product promotion, and 723,245 loan applicants.  
Bank partners operate under a hierarchical and viral growth model, which allows them to recruit downstream partners. Although this model facilitates the rapid expansion of lending operations, it also increases the risk propagation of loan fraud.
\textbf{Shareholding:} 
We collect data on the top 10 shareholders and their shareholding relationships for listed companies across 39 quarters, spanning from March 31, 2013, to September 30, 2022. The data is obtained from the China Stock Market \& Accounting Research (CSMAR) database (https://data.csmar.com/) and is used to analyze risks associated with large-scale stock liquidations by major shareholders, which can trigger adverse market reactions and lead to heightened price volatility.
%
%
\textbf{Bitcoin-OTC}~\cite{kumar2016edge}: It is a "who-trusts-whom" network of Bitcoin users trading on the platform \url{http://www.bitcoin-otc.com}. This dataset can be utilized for predicting the polarity of each rating and forecasting whether a user will rate another in the subsequent time step.
\textbf{Bitcoin-Alpha}~\cite{kumar2018rev2}: It is constructed similarly to Bitcoin-OTC, but the users and ratings originate from a different trading platform, \url{http://www.btc-alpha.com}.

\subsection{Hyperparameter Settings}
All parameters can be fine-tuned using 5-fold cross-validation on a rolling basis. For the dynamic graph learning module, we set the embedding dimension and the number of layers to $64$ and $3$, respectively. For the risk recognizing module, we set the balance weight $\tau_3$ and the number of layers to $0.9$ and $3$, respectively. Additionally, the framework is trained using the Adam optimizer with a learning rate of 0.0001. We train the Bitcoin-Alpha and Bitcoin-OTC datasets for 200 epochs, and the remaining two datasets for 300 epochs.

To validate the effectiveness of the dynamic graph learning module in subsequent subsections, we also develop $\text{GraphShield}^\dagger$, where separable function-based spatial and temporal operations are replaced with the \texttt{GCN+GRU} framework. 
To assess the performance of the risk recognizing module, we introduce $\text{GraphShield}^\ddagger$, 
in which the balance weight $\tau_3$ is set to zero. 
This adjustment removes $C_\mathrm{unlabel}$, transforming the module into a supervised variant.

\subsection{Baselines}
To demonstrate the efficacy of our proposed risk recognition framework, we employ two categories of baseline methods for comparison:
(a) Static graph methods, including Node2vec~\cite{grover2016node2vec}, GCN~\cite{kipf2016semi}, and GAT~\cite{velivckovic2018graph}.
(b) Dynamic graph methods, such as GAT-Informer (a hybrid of GAT and Informer~\cite{zhou2021informer}), GAT-PatchTST (a combination of GAT and PatchTST~\cite{nie2022time}), along with RisQNet~\cite{ijcai2024p817}, EvolveGCN~\cite{pareja2020evolvegcn}, StrGNN~\cite{cai2021structural}, TADDY~\cite{liu2021anomaly}, and AddGraph~\cite{zheng2019addgraph}.


\subsection{Overall Comparison \& Ablation Study}
\begin{table}[h]
\setlength{\tabcolsep}{0.6mm}
\setlength{\abovecaptionskip}{0pt}
\setlength{\belowcaptionskip}{0pt}
\fontsize{7pt}{6pt}\selectfont

\begin{tabular}{cccc|ccc|ccc|ccc} 
\toprule
\multicolumn{13}{c}{(a) \textbf{Unlabeled Ratio = 0\%}}       \\
Datasets & \multicolumn{3}{c|}{Bank-Partner} & \multicolumn{3}{c|}{Shareholding} & \multicolumn{3}{c|}{Bitcoin-OTC} & \multicolumn{3}{c}{Bitcoin-Alpha}  \\
\midrule
Risk Ratio    & 1\%    & 5\%    & 10\% & 1\%    & 5\%    & 10\%  & 1\%    & 5\%    & 10\%   & 1\%    & 5\%    & 10\% \\
 \midrule
Node2vec       & .725 & .718 & .726 & .570 & .563 & .550 & .695 & .688 & .674 & .691 & .680 & .678 \\
GCN  & .740 & .733 & .741 & .608 & .613 & .620 & .757 & .749 & .762 & .752 & .745 & .753 \\
GAT  & .750 & .743 & .751 & .625 & .615 & .614 & .767 & .759 & .772 & .761 & .754 & .763 \\
\midrule
EvolveGCN      & .781 & .774 & .783 & .647 & .653 & .660 & .801 & .793 & .806 & .795 & .788 & .797 \\
GAT-Informer   & .812 & .805 & .814 & .672 & .669 & .669 & .832 & .823 & .837 & .826 & .818 & .828 \\
GAT-PatchTST   & .818 & .810 & .820 & .686 & .660 & .667 & .838 & .829 & .843 & .832 & .824 & .834 \\
RisQNet & .859 & .862 & .863 & .718 & .715 & .714 & .890 & .881 & .895 & .883 & .876 & .885 \\
StrGNN & .888 & .865 & .870 & .686 & .712 & .704 & .901 & .878 & .884 & .857 & .867 & .863 \\
TADDY  & .931 & .921 & .928 & .771 & .755 & .764 & .946 & .934 & .943 & .945 & .934 & .942 \\
AddGraph       & .823 & .834 & .846 & .703 & .693 & .689 & .835 & .846 & .859 & .859 & .840 & .850 \\
\midrule
$\text{GraphShield}^\dagger$     & .885 & .893 & .885 & .783 & .784 & .776 & .916 & .912 & .895 & .911 & .910 & .901 \\
\textbf{GraphShield} & \textbf{.942} & \textbf{.940} & \textbf{.952} & \textbf{.833} & \textbf{.825} & \textbf{.834} & \textbf{.974} & \textbf{.960} & \textbf{.963} & \textbf{.969} & \textbf{.958} & \textbf{.970} \\
\bottomrule
\toprule
\multicolumn{13}{c}{(b) \textbf{Unlabeled Ratio = 40\%}}      \\
Datasets & \multicolumn{3}{c|}{Bank-Partner} & \multicolumn{3}{c|}{Shareholding} & \multicolumn{3}{c|}{Bitcoin-OTC} & \multicolumn{3}{c}{Bitcoin-Alpha}  \\
\midrule
Risk Ratio & 1\%    & 5\%    & 10\% & 1\%    & 5\%    & 10\%  & 1\%    & 5\%    & 10\%   & 1\%    & 5\%    & 10\% \\
 \midrule
Node2vec     & .620 & .615 & .625 & .501 & .501 & .508 & .594 & .587 & .571 & .589 & .577 & .575  \\
GCN          & .640 & .633 & .640 & .507 & .509 & .517 & .656 & .644 & .658 & .650 & .642 & .651  \\
GAT          & .646 & .642 & .651 & .522 & .510 & .513 & .664 & .657 & .670 & .659 & .654 & .662  \\
\midrule
EvolveGCN    & .676 & .673 & .682 & .544 & .552 & .557 & .699 & .689 & .702 & .691 & .687 & .694  \\
GAT-Informer & .711 & .704 & .710 & .570 & .565 & .564 & .730 & .719 & .735 & .725 & .717 & .725  \\
GAT-PatchTST & .718 & .709 & .718 & .583 & .556 & .565 & .733 & .724 & .740 & .728 & .719 & .730  \\
RisQNet      & .767 & .761 & .768 & .615 & .611 & .609 & .788 & .780 & .792 & .781 & .772 & .782  \\
StrGNN       & .787 & .761 & .768 & .586 & .608 & .603 & .801 & .774 & .780 & .757 & .764 & .760  \\
TADDY        & .830 & .818 & .823 & .668 & .654 & .661 & .843 & .830 & .840 & .843 & .831 & .841  \\
AddGraph     & .721 & .733 & .742 & .598 & .590 & .586 & .732 & .741 & .757 & .756 & .739 & .746  \\
\midrule
$\text{GraphShield}^\dagger$       & .787 & .797 & .790 & .685 & .685 & .679 & .817 & .814 & .801 & .813 & .812 & .809 \\
$\text{GraphShield}^\ddagger$        & .770 & .789 & .790 & .672 & .678 & .679 & .799 & .805 & .801 & .796 & .804 & .819 \\
\textbf{GraphShield} & \textbf{.837} & \textbf{.839} & \textbf{.849} & \textbf{.730} & \textbf{.721} & \textbf{.730} & \textbf{.869} & \textbf{.857} & \textbf{.861} & \textbf{.865} & \textbf{.855} & \textbf{.870}  \\
\bottomrule
\toprule
\multicolumn{13}{c}{(c) \textbf{Unlabeled Ratio = 90\%}}      \\
Datasets & \multicolumn{3}{c|}{Bank-Partner} & \multicolumn{3}{c|}{Shareholding} & \multicolumn{3}{c|}{Bitcoin-OTC} & \multicolumn{3}{c}{Bitcoin-Alpha}  \\
\midrule
Risk Ratio   & 1\%    & 5\%    & 10\% & 1\%    & 5\%    & 10\%  & 1\%    & 5\%    & 10\%   & 1\%    & 5\%    & 10\% \\
 \midrule
Node2vec     & .513 & .505 & .513 & .501 & .526 & .524 & .517 & .508 & .511 & .507 & .507 & .520  \\
GCN          & .526 & .519 & .528 & .506 & .529 & .520 & .507 & .529 & .505 & .530 & .519 & .519  \\
GAT          & .537 & .530 & .540 & .504 & .529 & .503 & .525 & .528 & .522 & .510 & .502 & .526  \\
\midrule
EvolveGCN    & .569 & .562 & .572 & .505 & .512 & .518 & .521 & .512 & .500 & .518 & .517 & .514  \\
GAT-Informer & .599 & .593 & .602 & .500 & .520 & .526 & .618 & .608 & .627 & .614 & .605 & .616  \\
GAT-PatchTST & .607 & .600 & .608 & .501 & .521 & .507 & .625 & .616 & .631 & .617 & .612 & .621  \\
RisQNet      & .660 & .649 & .657 & .503 & .508 & .525 & .677 & .669 & .684 & .673 & .661 & .672  \\
StrGNN       & .678 & .655 & .658 & .505 & .517 & .508 & .689 & .667 & .669 & .646 & .653 & .650  \\
TADDY        & .720 & .708 & .714 & .556 & .542 & .549 & .732 & .720 & .732 & .732 & .722 & .730  \\
AddGraph     & .612 & .621 & .631 & .513 & .511 & .512 & .621 & .635 & .649 & .648 & .628 & .637  \\
\midrule
$\text{GraphShield}^\dagger$       & .694 & .679 & .710 & .586 & .581 & .580 & .730 & .701 & .713 & .704 & .716 & .713 \\
$\text{GraphShield}^\ddagger$        & .672 & .685 & .688 & .573 & .569 & .587 & .707 & .686 & .706 & .704 & .686 & .723 \\
\textbf{GraphShield}  & \textbf{.730} & \textbf{.729} & \textbf{.740} & \textbf{.623} & \textbf{.612} & \textbf{.624} & \textbf{.760} & \textbf{.746} & \textbf{.751} & \textbf{.757} & \textbf{.746} & \textbf{.759}  \\
\bottomrule
\end{tabular}
\centering
\caption{\textbf{Risk Detection Performance in Terms of AUC}. 
The percentages $1\%$, $5\%$, and $10\%$ indicate the proportions of risk. 
The percentages $0\%$, $40\%$, and $90\%$ indicate the proportions of unlabeled data containing potential financial risks. 
As the proportion rises, so does the difficulty in identifying these hidden financial risks.
}
\label{tab:risk-recognition}
\end{table}
Table~\ref{tab:risk-recognition} presents a comparison of risk recognition performance based on the average AUC across all test timestamps. 
From these results, we observe that: 
(a) The GAT model outperforms the GCN model, benefiting from the attention mechanism. This finding underscores the effectiveness of attention in handling complex relationships within graphs.
(b) Dynamic graph models surpass static graph models, emphasizing the significant role of temporal dynamics in accurately identifying risks. This highlights the necessity of incorporating time-evolving data for better predictive accuracy.
(c) The GraphShield framework consistently shows superior and robust performance, particularly as challenges such as limited labels and class imbalance intensify. In contrast, the performance of other models significantly declines. This robustness suggests that GraphShield effectively addresses the complexities arising from sparse and unevenly distributed data.
(d) Compared to models like $\text{GraphShield}^\dagger$, GAT-Informer, GAT-PatchTST, RisQNet, StrGNN, and AddGraph, the GraphShield framework excels due to its dynamic graph learning module. This module enhances the integration of spatial and temporal information, leading to better risk detection.
(e) As the proportion of unlabeled data increases from $40\%$ to $90\%$, GraphShield maintains an average performance advantage of over $6\%$ compared to $\text{GraphShield}^\ddagger$. This advantage is largely due to our approach's reliance on clustering and the tailed distribution of risks, which provides greater adaptability.


\subsection{Hyperparameter Sensitivity Analysis}
In this section, we investigate the influence of hyperparameters on GraphShield, focusing on the embedding dimension and number of layers in the dynamic graph learning module, the balance weight $\tau_3$ of the loss function, and the number of layers in the semi-supervised risk detecting module. We conduct experiments on two datasets: Bitcoin-Alpha and Bitcoin-OTC. Throughout these experiments, we keep all other parameters at their default settings and evaluate performance in an environment with a $10\%$ risk proportion and $100\%$ label proportion.

\begin{figure}[h]
    \setlength{\abovecaptionskip}{0pt}
    \setlength{\belowcaptionskip}{0pt}
    \centering

    \begin{subfigure}[b]{0.23\textwidth}
        \centering
        \includegraphics[width=\linewidth]{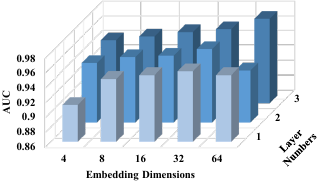}
    \caption{Bitcoin-OTC}
    \label{fig:Bitcoin-OTC}
    \end{subfigure}
    \begin{subfigure}[b]{0.24\textwidth}
        \centering
        \includegraphics[width=\linewidth]{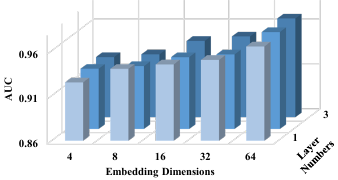}
        \caption{Bitcoin-Alpha}
        \label{fig:Bitcoin-Alpha}
    \end{subfigure}
    \caption{\textbf{Sensitivity of Embedding Dimension and Layer Number in Dynamic Graph Learning Module}.}
    \label{fig:sentivity}
\end{figure}
\vspace{-4mm}
\begin{figure}[h]
    \setlength{\abovecaptionskip}{0pt}
    \setlength{\belowcaptionskip}{0pt}
    \centering

    \begin{subfigure}[b]{0.23\textwidth}
        \centering
        \includegraphics[width=\linewidth]{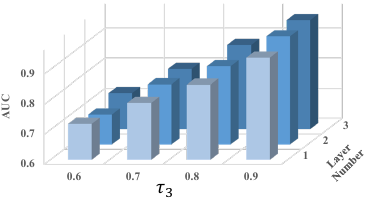}
    \caption{Bitcoin-OTC}
    \label{fig:Bitcoin-OTC2}
    \end{subfigure}
    \begin{subfigure}[b]{0.24\textwidth}
        \centering
        \includegraphics[width=\linewidth]{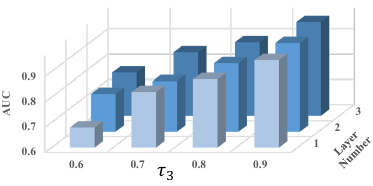}
        \caption{Bitcoin-Alpha}
        \label{fig:Bitcoin-Alpha2}
    \end{subfigure}
    \caption{\textbf{Sensitivity of Balance Weight $\tau_3$ and Layer Number in Semi-supervised Risk Detecting Module}.}
    \label{fig:sentivity2}
\end{figure}

\begin{figure*}[h]
    \setlength{\abovecaptionskip}{0pt}
    \setlength{\belowcaptionskip}{0pt}
    \centering
    \begin{subfigure}[b]{0.7\textwidth}
        \centering
        \includegraphics[width=\linewidth]{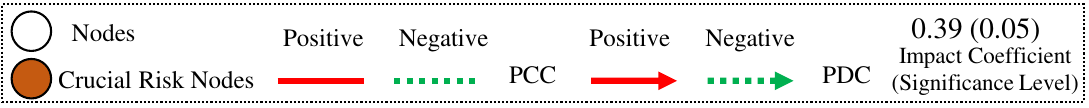}
    \end{subfigure}
    
    \begin{subfigure}[b]{0.25\textwidth}
        \centering
        \includegraphics[width=\linewidth]{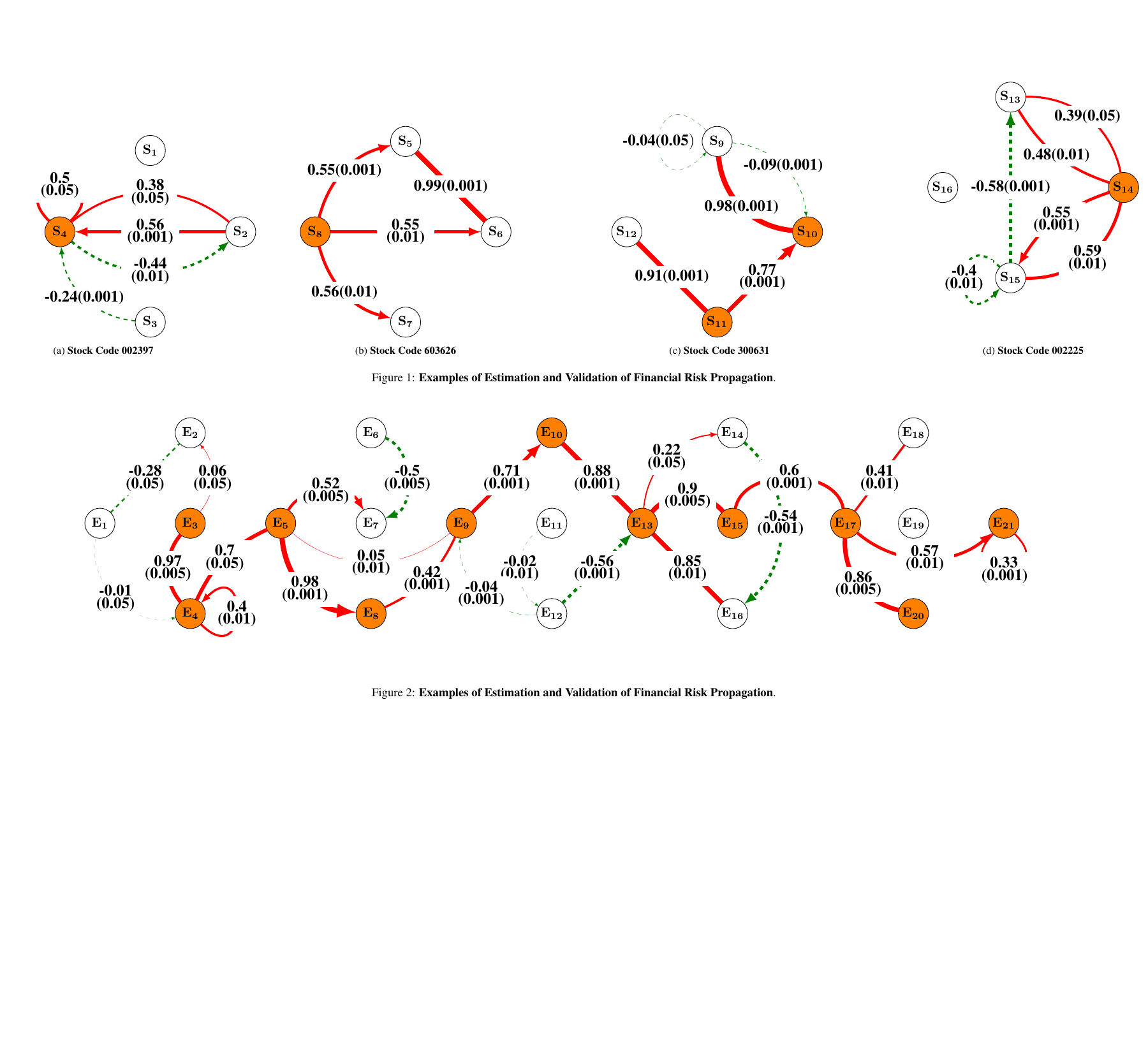}
    \caption{Crucial Risk Node $\mathrm{S}_4$}
    \label{fig:code002397}
    \end{subfigure}
    \begin{subfigure}[b]{0.23\textwidth}
        \centering
        \includegraphics[width=\linewidth]{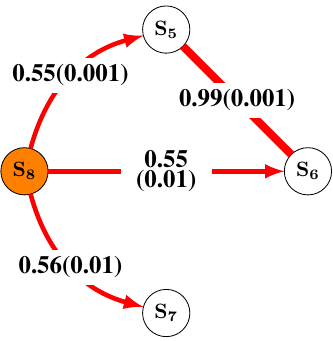}
        \caption{Crucial Risk Node $\mathrm{S}_8$}
        \label{fig:code603626}
    \end{subfigure}
%
    \begin{subfigure}[b]{0.25\textwidth}
        \centering
        \includegraphics[width=\linewidth]{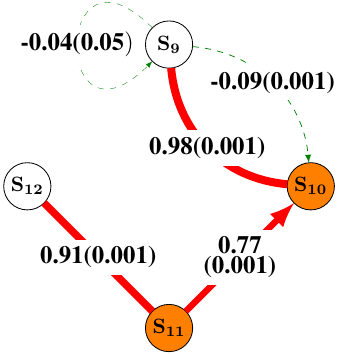}
        \caption{Crucial Risk Node $\mathrm{S}_{10}$ and $\mathrm{S}_{11}$}
        \label{fig:code300631}
    \end{subfigure}
    %
    \begin{subfigure}[b]{0.23\textwidth}
        \centering
        \includegraphics[width=\linewidth]{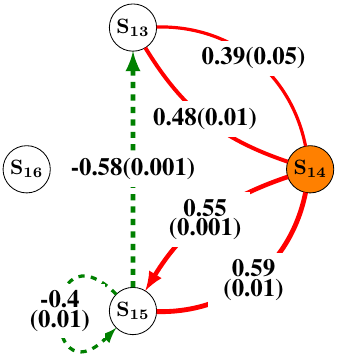}
        \caption{Crucial Risk Node $\mathrm{S}_{14}$}
        \label{fig:code002225}
    \end{subfigure}
    \caption{\textbf{Examples of Estimation and Verification of Impact Effects in Stock Sell-off Risk Propagation}. Such examples are extracted from the Shareholding dataset, which serves as the basis for analyzing risk propagation mechanisms.
    PCC is used to quantify bidirectional effects of risk propagation in the spatial domain, 
    whereas PDC focuses on unidirectional effects in the temporal domain.
    PCC and PDC values with $p\text{-value} \geq 0.05$ are considered statistically insignificant and are thus excluded from further analysis.
    The shareholder nodes $\mathrm{S}_{4}$, $\mathrm{S}_{8}$, $\mathrm{S}_{10}$, $\mathrm{S}_{11}$, and $\mathrm{S}_{14}$ have been identified as risk nodes by our proposed approach and are highlighted in orange. Notably, these nodes align with the actual labels and play a critical role in risk propagation, warranting special attention.
    }
    \label{fig:selloff}
\end{figure*}

\begin{figure*}[h]
    \setlength{\abovecaptionskip}{0pt}
    \setlength{\belowcaptionskip}{0pt}
    \centering
    \includegraphics[width=\linewidth]{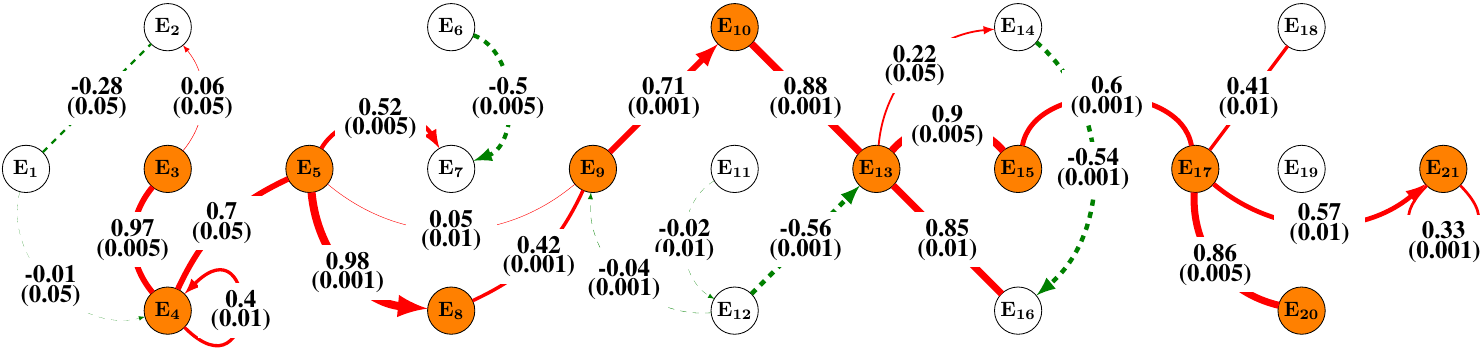}
    \caption{\textbf{Examples of Estimation and Verification of Impact Effects in Loan-Fraud Risk Propagation}. These examples are drawn from the Bank-Partner dataset. Partner and customer nodes, such as $\mathrm{E}_3$ and $\mathrm{E}_4$, have been identified as risk nodes by our proposed approach and are highlighted in orange. Notably, these nodes correspond to the actual labels and play a critical role in risk propagation, requiring special attention.}
    \label{fig:loandefault}
\end{figure*}

In the dynamic graph learning module, we explore embedding dimensions ranging from $\{4, 8, 16, 32, 64\}$ and layer numbers from $\{1, 2, 3\}$. The sensitivity of these hyper-parameters is depicted in Fig.~\ref{fig:sentivity}. In the semi-supervised risk recognizing module, we adjuste the balance weight $\tau_3$ within $\{0.6, 0.7, 0.8, 0.9\}$ and the layer numbers from $\{1, 2, 3\}$. The impact of these settings is shown in Fig.~\ref{fig:sentivity2}. These figures clearly demonstrate that both embedding dimensions and layer numbers significantly enhance AUC, with the optimal settings being an embedding dimension of 64 and a layer number of 3. Notably, the balance weight $\tau_3$ exhibits the most substantial improvement in AUC, increasing from approximately 0.7 to 0.9, underscoring the importance of the supervised constraint $C_\mathrm{label}$ in boosting model performance.

\section{Visualization Analysis of Risk Propagation}
In this section, we utilize the visualization analysis tool to conduct the analysis of stock sell-off risk propagation on the Shareholding dataset.
Here, we select dynamic subgraphs featuring the top-4 shareholders (that is, $n=4$) from June 2013 to June 2022 from the Shareholding dataset. We then estimate and validate the impact of stock sell-off risk propagation, using the optimal lag of $I=1$ as determined by the Akaike Information Criterion.
The experimental results are displayed in Fig.~\ref{fig:selloff}, giving four examples of estimation and verification of impact effects 
via selecting four representative stocks:
002397 (MENDALE), 603626 (KERSEN), 300631 (JIUWUHI-TECH), and 002225 (PRCO). 
From this figure, we can observe that: 
(a) The predicted risk labels align the ground-truth labels, highlighted in orange.
(b) When a shareholder significantly reduces their holdings, the likelihood of this action is influenced by the divestment tendencies of associated shareholders in the current period as well as divestment patterns from the previous period. (c) Shareholder nodes $\mathrm{S}_{3}$, $\mathrm{S}_{8}$, $\mathrm{S}_{10}$, $\mathrm{S}_{11}$, and $\mathrm{S}_{14}$ play major roles in risk propagation and should be closely monitored.

Additionally, we perform a similar analysis of loan-fraud default risk propagation using the Bank-Partner dataset. As shown in Fig.~\ref{fig:loandefault}, key nodes marked in red, such as $\mathrm{E}_3$ and $\mathrm{E}_4$, play a crucial role in the propagation chain.

\section{Conclusion}
This study introduces GraphShield, an innovative and effective dynamic graph learning model designed to safeguard financial stability against risk propagation. This approach achieves three key functionalities: (a) enhancing information learning across temporal and spatial domains, (b) improving hidden risk recognition, and (c) visualizing and analyzing risk propagation process.
Various experiments on two real-world datasets and two public datasets highlights the strong performance of our approach.
In addition, our approach has already been successfully deployed at an Internet commercial bank in Sichuan, where it is already demonstrating tangible impact. 

Furthermore, We are actively enhancing GraphShield functionality, with plans to deploy it in critical sectors such as supply chain finance and banking risk management. This is expected to significantly boost financial stability and contribute to sustainable economic development.
\newpage
\bibliographystyle{named}
\bibliography{ref}

\newpage
\appendix
\section{Likelihood Ratio Test}
This section presents the likelihood ratio test for the estimated impact coefficients obtained through the visualization analysis tool.
\begin{equation}
    \begin{aligned}
       &\text{Null Hypothesis}~(H_0): \bO^{(\ell)}_{i,j}=0~(\text{or}~\bOmega_{i,j})=0),\\
       &\text{Alternative Hypothesis}~(H_1): \bO^{(\ell)}_{i,j}\neq0~(\text{or}~\bOmega_{i,j})\neq0).
    \end{aligned}
\end{equation}
Based on the above null hypothesis, the likelihood ratio test statistic is defined as, 
\begin{equation}
    \begin{aligned}
       \text{LRT} = -2(\ln L_R - \ln L_F),
    \end{aligned}
\end{equation}
where $\ln L_F$ and $\ln L_R$ denote the log-likelihood values for the cases where $\bO^{(\ell)}_{i,j} \neq 0$ (full model) and $\bO^{(\ell)}_{i,j} = 0$ (reduced model), respectively. 

Under large-sample conditions, the Maximum Likelihood Estimation (MLE) is both consistent and asymptotically normal.
This means that the MLE converges to the true parameter values as the sample size increases, and its distribution approaches a normal distribution. This property forms the basis for Wilks' theorem, which states that under the null hypothesis, the LRT statistic asymptotically follows a chi-square distribution. Specifically, if the full and reduced models have $p_F$ and $p_R$ parameters, respectively, the difference in the number of parameters, $\Delta p = p_F - p_R$, determines the degrees of freedom for the chi-square distribution. Thus, the asymptotic distribution of the LRT statistic is given by,

\begin{equation}
    \begin{aligned}
    \text{LRT} \sim \chi^2(\Delta p).
    \end{aligned}
\end{equation}

The derivation of this result relies on the properties of the log-likelihood function, which can be approximated as a quadratic function near the MLE. Using a Taylor expansion around the MLE, the log-likelihood function can be expressed as,

\begin{equation}
    \begin{aligned}
    \ln L(\theta) \approx \ln L(\hat{\theta}) + \frac{1}{2}(\theta - \hat{\theta})^\top I(\hat{\theta})(\theta - \hat{\theta}),
    \end{aligned}
\end{equation}
where $I(\hat{\theta})$ is the Fisher information matrix. Given that the log-likelihood function is approximately quadratic and the MLE is asymptotically normal, it follows that

\begin{equation}
    \begin{aligned}
-2(\ln L_R - \ln L_F) \approx (\hat{\theta}_F - \hat{\theta}_R)^\top I(\hat{\theta}_R)(\hat{\theta}_F - \hat{\theta}_R).
    \end{aligned}
\end{equation}
Under large-sample conditions, this quantity asymptotically follows a chi-square distribution with $\Delta p$ degrees of freedom.



\section{Computational Complexity}
This section analyzes the computational complexity of the proposed GraphShield framework. 
In the dynamic graph learning module, the computational complexity is primarily determined by the spatial operation $\mathcal{O}(T\bar{m}H)$ and temporal operation $\mathcal{O}(T\bar{n}H)$, where $T$ represents the total number of timestamps, $\bar{m}$ and $\bar{n}$ denote the average number of edges and nodes per graph snapshot, respectively, and $H$ is the total number of attention heads. Consequently, for a $L$-layer dynamic graph learning module, the overall computational complexity is $\mathcal{O}(LTH(\bar{m}+\bar{n}))$.
In the risk recognizing module, the primary computational costs stem from the fully-connected network ($\mathcal{O}(dL)$) and covariance calculation ($\mathcal{O}(d^2)$), with $d$ representing the embedding dimension.
In the visualization analysis tool, the computational complexity is dominated by the calculation of $\bO$ and $\bOmega$, which is $\mathcal{O}(T I n^2 )$.

\section{Pseudocode of Our Framework}

	
        
\vspace{-4mm}
\begin{algorithm}[h!]
    \small
    \caption{Risk Recognition}
    \label{algo:recognition}
    \LinesNumbered
    \KwIn{Financial dynamic graphs $\mathcal{G} = \{\mathcal{G}^{\mathsf{S}}, \mathcal{G}^{\mathsf{T}}\}$, maximum training epoch $E$, maximum timestamp $T$.}
    \KwOut{Risk probability $\gamma$ for each node.}
    \BlankLine
    Randomly initialize the parameters of the dynamic graph learning module and the risk recognition module\;
    
    \For{epoch $e = 1$ to $E$}{
        \For{timestamp $t = 1$ to $T$}{
            Compute the spatial representations of all nodes in $\mathcal{G}^\mathsf{S}_t$ using Eqs.~(1) and (2)\;
        }
        
        \For{each node $u \in \mathcal{A}$}{
            Add rotary encoding to embeddings of $u$ for $t=1$ to $T$ via Eqs.~(3) and (4)\;
            Compute the temporal representations of $u$ for the next layer using Eq.~(5)\;  
        }
        
        Obtain the final node embeddings $\mathcal{Z}$\; 
        Compute the risk probabilities $\gamma$ through a fully-connected network using Eqs.~(6) and (7)\;
        Calculate the estimated expectation $\widehat{\bmu}_k$, component probability $\hat{\pi}_k$, and covariance $\widehat{\bSigma}_k$ using Eq.~(8)\;
        Calculate the final loss function in Eq.~(9)\;
        Perform backpropagation and update the parameters\;
    }
    \BlankLine
\end{algorithm}

\vspace{-8mm}

\begin{algorithm}[h!]
    \small
    \caption{Visualization of Risk Propagation}
    \label{algo:visualization}
    \LinesNumbered
    \KwIn{Financial dynamic graphs $\mathcal{G} = \{\mathcal{G}^{\mathsf{S}}, \mathcal{G}^{\mathsf{T}}\}$, maximum training epoch $E$.}
    \BlankLine
    \For{lag order $I \in \{1, 2, 3\}$}{
        Randomly initialize the model parameters\;
        \For{epoch $e = 1$ to $E$}{
            Compute the model output using Eq.~(10)\;
            Calculate the loss function using Eq.~(11)\;
            Perform backpropagation and update the parameters\;
        }
    }
    Determine the optimal lag order based on AIC\;
    \For{each element $o \in \bO$ or $\bOmega$}{
        Construct the likelihood ratio test statistic using Eq.~(15)\;
        Calculate the $p$-value\;
        \If{$p$-value $\geq 0.05$}{
            Remove element $o$\;
        }
    }
    \BlankLine
\end{algorithm}

\end{document}